# Bottom Contact Metal Oxide Interface Modification Improving the Efficiency of Organic Light Emitting Diodes

Sergey M. Pozov [1], Apostolos Ioakeimidis [1], Ioannis T. Papadas [1], Chen Sun [2], Alexandra Z. Chrusou [1], Donal D. C. Bradley [2,3] and Stelios A. Choulis [1,*]

[1] Department of Mechanical Engineering and Materials Science and Engineering, Molecular Electronics and Photonics Research Unit, Cyprus University of Technology, Limassol 3603, Cyprus; sergey.pozov@cut.ac.cy (S.M.P.); a.ioakeimidis@cut.ac.cy (A.I.); ioannis.papadas@cut.ac.cy (I.T.P.); az.chrusou@cut.ac.cy (A.Z.C.)

[2] Department of Physics, Clarendon Laboratory, University of Oxford, Parks Rd, Oxford OX1 3PU, UK; chen.sun@physics.ox.ac.uk (C.S.); donal.bradley@kaust.edu.sa (D.D.C.B.)

[3] Physical Science and Engineering Division, King Abdullah University of Science and Technology (KAUST), Thuwal 23955-6900, Saudi Arabia

* Correspondence: stelios.choulis@cut.ac.cy

**Abstract:** The performance of solution-processed organic light emitting diodes (OLEDs) is often limited by non-uniform contacts. In this work, we introduce Ni-containing solution-processed metal oxide (MO) interfacial layers inserted between indium tin oxide (ITO) and poly(3,4-ethylenedioxythiophene):poly(styrene sulfonate) (PEDOT:PSS) to improve the bottom electrode contact for OLEDs using the poly(p-phenylene vinylene) (PPV) derivative Super-Yellow (SY) as an emission layer. For ITO/Ni-containing MO/PEDOT:PSS bottom electrode structures we show enhanced wetting properties that result in an improved OLED device efficiency. Best performance is achieved using a Cu-Li co-doped spinel nickel cobaltite [(Cu-Li):NiCo$_2$O$_4$], for which the current efficiency and luminous efficacy of SY OLEDs increased, respectively, by 12% and 11% from the values obtained for standard devices without a Ni-containing MO interface modification between ITO and PEDOT:PSS. The enhanced performance was attributed to the improved morphology of PEDOT:PSS, which consequently increased the hole injection capability of the optimized ITO/(Cu-Li):NiCo$_2$O$_4$/PEDOT:PSS electrode.

**Keywords:** interfaces; electrodes; wetting properties; metal-oxides; hole injection; organic light emitting diodes

## 1. Introduction

Since the discovery of polymer light emitting diodes [1], important progress has been accomplished, achieving highly efficient, stable, and flexible solution-processed organic light emitting diode (OLED) devices [2,3]. The electroluminescent (EL) efficiency of such devices depends on many factors, including the active layer quantum yield, the recombination and transport properties as well as the device architecture and engineering of electrodes and interfaces [4,5]. High quality interfaces are very important in OLED applications to achieve efficient charge injection and avoid low dielectric breakdown voltage and dark spot formation [6,7]. The injection efficiency of holes from the most widely used indium tin oxide (ITO)/Poly(3,4-ethylenedioxythiophene):poly(styrene sulfonate) (PEDOT:PSS) hole injection electrode (HIE) [8], strongly depends on the energy level alignment between the PEDOT:PSS/organic active layer [9–14] and the interfacial morphology between the various layers (e.g., ITO, PEDOT:PSS, organic active layer) [15,16]. To achieve high quality ITO/PEDOT:PSS interface and efficient hole injection, two main methods are usually applied in conventional solution-processed OLEDs. First, the addition of specific surfactants or solvents into PEDOT:PSS solution [17,18], and second, the modification of ITO surface with treatments such as plasma or UV-ozone [19,20], where both methods are designed to improve the wettability of ITO and therefore the quality of the deposited film. Additionally, the pre-treatment of ITO surface results in significant increase in the WF from 4.5 eV up to 4.7 eV and favors the efficiency of OLED devices [21,22].

In this work, we introduce Ni-containing solution-processed metal oxide (MO) interfacial layers between ITO and PEDOT:PSS, to improve bottom contact properties. Metal oxides synthesized in air usually present surfaces rich in hydroxyl groups, which can be exploited to increase the wettability of PEDOT:PSS on hydrophobic ITO surfaces [23]. Previous works have shown that the configuration of ITO/MO/PEDOT:PSS can improve the device efficiency in phosphorescent and fluorescent OLEDs, as well as in quantum dot LEDs, using some of the well-known MOs, such as $NiO_x$, $WO_x$, and $MoO_x$. The improvement in device efficiency were interpreted in terms of energetic alignment with stair-like hole injection [24,25] or improved electrical conductivity of PEDOT:PSS due to doping by the used metal oxide [26]. Recently, Chen and his co-workers published a work incorporating double stack hole injection ITO/$NiO_x$/PEDOT:PSS electrode in phosphorescent-LED, increasing the device efficiency. The increase in device efficiency was attributed to reduced exciton quenching of the $NiO_x$ surface, improved electron blocking of the proposed electrode, and the increased hole injection due to energy pinning effect of $NiO_x$ and PEDOT:PSS [27].

In this report, we propose that the use of suitable Ni-containing solution-processed (SCS) MO interfacial layer between ITO and PEDOT:PSS can be used as a method to improve the properties of the bottom electrode (ITO/MO/PEDOT:PSS). Un-doped spinel nickel cobaltite ($NiCo_2O_4$) and co-doped with 3% Cu and 2% Li [(Cu-Li):$NiCo_2O_4$] [28,29] as well as Cu:$NiO_x$ and $CuO_x$ [30–32] were used as MOs investigated in the current work. We show that Ni-containing MOs interface modification ITO/MO/PEDOT:PSS improved device performance in normal structure fluorescent super-yellow (SY) OLED devices, in comparison to reference ITO/PEDOT:PSS bottom electrode. Contact angle measurements demonstrated improved wetting of Ni-containing MOs (undoped-$NiCo_2O_4$, co-doped (Cu-Li):$NiCo_2O_4$ and Cu:$NiO_x$) in comparison with the poor wetting of $CuO_x$. For the best performing co-doped (Cu-Li):$NiCo_2O_4$ interface modification, the improved wetting properties resulted in smooth and homogeneous PEDOT:PSS layer. Additionally, photocurrent mapping measurements of the corresponding OLED devices, provided evidence that the proposed interface modification improved the performance of the bottom electrode, resulting in improved hole injection properties which in turns increased both current efficiency and power efficacy of super-yellow OLED device.

## 2. Materials and Methods

*2.1. Materials*

Pre-patterned glass-indium tin oxide (ITO) substrates (sheet resistance 4 $\Omega.sq^{-1}$) were purchased from Psiotec Ltd. Hole injection layer, Clevios P VP AI4083 low conductivity grade (3,4-ethylenedioxylthiophene): poly(styrene sulfonate) (PEDOT:PSS) acquired from Ossila LTD, and was filtered through a 0.22 μm polyvinylidene difluoride (PVDF) filter prior processing. Solution-processable poly (p-phenylene vinylene) PPV-based polymer, PDY-132 Livilux® Super-Yellow (SY) was purchased from Merck and used as a light emissive polymer layer. SY was dissolved inside an inert glovebox atmosphere using toluene solvent and a concentration of 6.5 mg/mL. All the other chemicals were purchased from Sigma-Aldrich.

*2.2. Combustion Synthesis and Deposition of Metal Oxides*

For the combustion synthesis and deposition of $NiCo_2O_4$ nanoparticles (NPs), (Cu-Li):$NiCo_2O_4$ NPs, and Cu:$NiO_x$ more details can be found elsewhere [28,29,33]. $CuO_x$ thin films were prepared by spin coating using 1 mmol Cu(NO$_3$)$_2$·3H$_2$O and 0.2 mmol of acetylacetone as fuel in 10 mL 2-methoxyethanol as starting materials to form a stable solution in deep green. The obtained solution was stirred for 1 h at room temperature. Then, the final precursor solution was spin-coated with a speed of 3000 rpm for 40 s. After spin-coating, the substrates were annealed at 300 °C on a hot plate for 1 h in air.

*2.3. Device Fabrication*

2.3.1. OLED devices

Organic light emitting diodes with normal structure and different hole injection layers (HIL) ITO/HIL/SY/LiF/Al were fabricated as follows. ITO were subsequently cleaned in acetone and isopropanol under sonication for 10 min, and then dried and exposed to uv-ozone treatment for 5 min. The ~40 nm PEDOT:PSS films were coated using static spin coating (Delta 6RC-Suss MicroTec, Garching, Germany) at 4000 rpm for 30 sec, followed by annealing at 140 °C for 20 min, under ambient atmosphere. Metal oxides were deposited using doctor blade (Erichsen, Hemer, Germany) or spin coater technique and their thickness was controlled through the concentration of the used solution. Super-yellow solution was spin coated at 6000 rpm for 30 s resulting in a thickness of ~80 nm, and simultaneously annealed on a hotplate at 50 °C for 20 min. Both spin coating and annealing were performed, under inert glovebox atmosphere. To complete the normal-structured stack, 1 nm (0.2 A/s) of lithium fluoride (LiF) and 100 nm (2 A/s) of aluminum (Al) were thermally evaporated in a vacuum chamber (Angstrom Engineering, Kitchener, Canada) at a base pressure of ~2 × $10^{-6}$ mbar through a shadow mask resulting in an active area of 9 $mm^2$. As a final step, all the devices were encapsulated using a glass slide and a UV–curable encapsulation epoxy from Ossila Ltd. (Sheffield, England)

2.3.2. Single-Carrier OLED Device

Hole only devices (HODs) configuration were fabricated using the following device structure ITO/HIL/SY/$MoO_3$/Au, where the same processing steps and materials were used as mentioned in OLED devices except the top electrode. $MoO_3$ HIL and Au electrode were thermally evaporated to form 10 nm (0.2 A/s) and 100 nm (2 A/s) layers, respectively.

*2.4. Characterization*

For film characterizations Schimadzu, UV-2700 UV–vis spectrophotometer was used to measure the transmittance spectra of the investigated materials on ITO substrates. Jasco FP-8300 spectrofluorometer with a corresponding 400 nm excitation wavelength was used to define the PL spectra of SY light emitting polymer (LEP) on various HILs on quartz substrates. AFM images were obtained using a Nanosurf easyScan 2 controller under the tapping mode. The thickness of the films was measured with a Veeco Dektak 150 profilometer. Eventually, complete devices were characterized using current density–voltage-luminance characteristics (JVL), which were obtained using a Botest LIV Functionality Test System with a calibrated silicon photodiode sensor (spectral sensitivity 350–730 nm and responsivity 60 nA/lux). All the characterizations were performed in ambient conditions.

**3. Results and Discussion**

The electrode performance of OLEDs strongly depended on the electrical conductivity, the work function, energy levels of electronic materials, and their wetting properties. Using the sessile drop technique, we investigated the wetting properties of the PEDOT:PSS (AI4083) on ITO and ITO/(Cu-Li):$NiCo_2O_4$ surfaces, as presented in Figure 1. The high static contact angle (CA) of un-modified hydrophobic ITO surface, reduced from 76.7° to 21.5° when the surface was treated with $UVO_3$, and down to 6.2° when (Cu-Li):$NiCo_2O_4$ interlayer was incorporated. Defining the $UVO_3$ treated ITO surface as our reference PEDOT:PSS coating condition, significantly improved wetting was achieved when (Cu-Li):$NiCo_2O_4$ interfacial incorporated between ITO and PEDOT:PSS. The excellent wetting of (Cu-Li):$NiCo_2O_4$ may be attributed to the combination of the surface chemistry and the low surface roughness of the material (3.0 nm) as shown in Figure 2a. Both surface chemistry and roughness strongly depend on the fuel regent used during the nanoparticle (NP) synthesis [29].

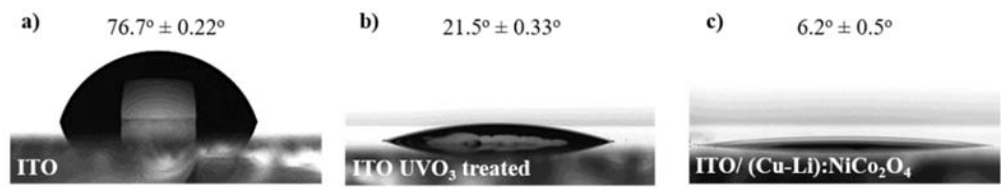

**Figure 1.** Static contact angle (CA) of deionized water droplet images on different substrates, (**a**) unmodified ITO, (**b**) ITO exposed to 5 min UVO$_3$ surface treatment, (**c**) ITO/(Cu-Li):NiCo$_2$O$_4$.

Additional investigation of the wetting properties of solution combustion synthesized (SCS) metal oxides (MOs) was performed using Cu:NiO$_x$ and CuO$_x$, as presented in Figure S1 [Supporting Information]. Similarly, improved wetting was achieved using Cu:NiO$_x$ interface modification, with CA of ~17°. On the other hand, CuO$_x$ did not show the same positive effects resulting to the highest CA of 47° among all the investigated metal oxides (MOs). The origin of the improved wetting properties in the case of Ni-based SCS metal oxides might be related to Ni/oxygen vacancies effects that influence the wetting properties. H$_2$O as coordinate hydroxyl group (OH-) is adsorbed on the oxygen vacancy site and bonds with the lattice of MO resulting the improvement of the surface wettability. Higher amounts of oxygen vacancies correspond to more OH content in the surface of the metal oxide and favorable wettability [34]. The different wettability behavior of SCS based CuO$_x$ compared to SCS Ni-containing MOs (Cu:NiO$_x$, un-doped NiCo$_2$O$_4$ and co-doped (Cu-Li):NiCo$_2$O$_4$)) could be attributed to the difference in those oxygen vacancies. In the case of Ni-based MOs, the oxidation states of nonstoichiometric NiO$_x$ can be attributed to the creation of surface Ni$^{3+}$ [35]. The oxidation of Ni$^{2+}$ to Ni$^{3+}$ during the process of synthesis, introduces positive holes in the MO crystal lattice and forms charged chemisorbed oxygen species on NiO$_x$. This can induce defects and increases the ionic vacancies in Ni-based MOs, which might be correlated with better electrical conductivity and wetting properties compare to CuO$_x$. Despite our initial studies, further investigations on the underline mechanism will be investigated in more detail in future publications.

Among the examined Ni-containing SCS metal oxides, we focused our studies on (Cu-Li):NiCo$_2$O$_4$ due to its superior wetting properties. Optical and morphological investigations of ITO/(Cu-Li):NiCo$_2$O$_4$, and ITO/(Cu-Li):NiCo$_2$O$_4$/PEDOT:PSS were performed and compared with ITO/PEDOT:PSS, which is considered as a reference bottom electrode. Figure 2a shows the surface topography images of ITO/(Cu-Li):NiCo$_2$O$_4$, ITO/PEDOT:PSS and ITO/(Cu-Li):NiCo$_2$O$_4$/PEDOT:PSS, obtained using atomic force microscopy (AFM). All the films showed a relatively smooth, pinhole-free, and homogeneous surface. ITO/(Cu-Li):NiCo$_2$O$_4$/PEDOT:PSS hole injection layer, exhibited root-mean-square (RMS) roughness and peak-to-valley of 1.9 nm and 13 nm, respectively. In contrast, ITO/PEDOT:PSS presented 2.4 nm and 16.5 nm, while ITO/(Cu-Li):NiCo$_2$O$_4$ 3.0 nm and 23.3 nm RMS and peak-to-valley, respectively. In complete agreement with the wetting results presented above, the proposed hydrophilic (Cu-Li):NiCo$_2$O$_4$ interface modification improved PEDOT:PSS surface topography.

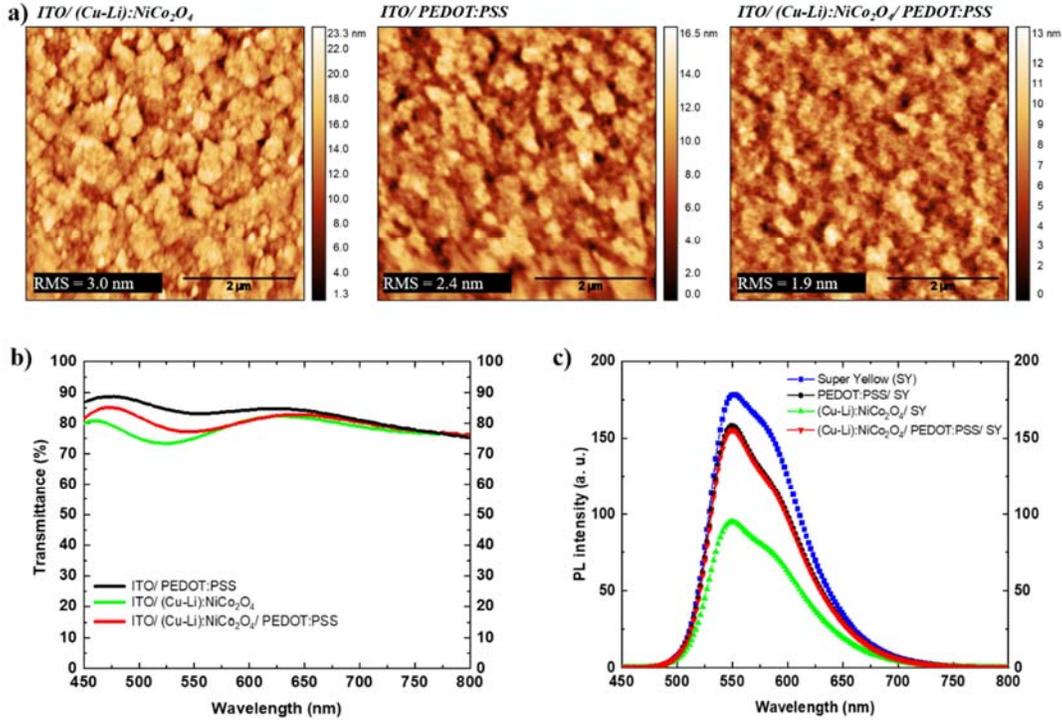

**Figure 2.** Optical and morphological characterization of the investigated hole injection electrodes, ITO/(Cu-Li):NiCo$_2$O$_4$, ITO/PEDOT:PSS and ITO/(Cu-Li):NiCo$_2$O$_4$/PEDOT:PSS, (**a**) AFM 5 × 5 μm images, (**b**) optical transmission spectra and (**c**) steady-state room temperature photoluminescence intensities of super-yellow (~80 nm) and HIL on quartz substrates. In all the cases, thicknesses were 40 nm for PEDOT:PSS and 15 nm for (Cu-Li):NiCo$_2$O$_4$.

Optical properties were studied using UV-visible (UV-Vis) optical transmission and steady-state photoluminescence (PL) spectra as shown in Figure 2b,c respectively. The optical transmittance measurements were limited in the 500–650 nm wavelength range, which is correlated to the PL and electroluminescence (EL) spectra of super-yellow light emitting polymer. In that range, the highest transmittance was recorded in ITO/PEDOT:PSS reference bottom electrode, with an average transmittance of ~84%. The incorporation of a (Cu-Li):NiCo$_2$O$_4$ SCS metal oxide interfacial layer slightly reduced the average transmittance down to ~80% for the ITO/(Cu-Li):NiCo$_2$O$_4$/PEDOT:PSS, and ~78% for the ITO/(Cu-Li):NiCo$_2$O$_4$. The differences in optical transmittance can be attributed to the lower optical bandgap of (Cu-Li):NiCo$_2$O$_4$ and thin film interference phenomenon. To study the exciton quenching phenomena, steady-state room temperature PL spectra of SY (80 nm) layer fabricated on quartz, quartz/(Cu-Li):NiCo$_2$O$_4$ (15 nm) and quartz/(Cu-Li):NiCo$_2$O$_4$/PEDOT:PSS (40 nm) were obtained. Despite the similar PL shapes, (Figure 2c) the lowest PL intensity was measured for the quartz/(Cu-Li):NiCo$_2$O$_4$/SY sample, indicating that significant exciton quenching occurred at the (Cu-Li):NiCo$_2$O$_4$/SY interface. Exciton quenching can be attributed to the lack of charge blocking ability of the bottom electrode when is in contact with emission layer [36]. The low energetic barrier formed between LUMOs of (Cu-Li):NiCo$_2$O$_4$ (~3.0 eV) and SY (~2.6 eV) creates a path for electrons and therefore reduces its electron blocking ability (see Figure 3d) resulting in increased exciton dissociation. Furthermore, exciton quenching may also arise due to surface defects of the MOs as well as the quality of the morphology of the deposited polymer film [37,38]. On the other hand, the PL spectra of quartz/(Cu-Li):NiCo$_2$O$_4$/PEDOT:PSS/SY and quartz/PEDOT:PSS/SY are almost identical in shape and intensity. Therefore, we infer that PEDOT:PSS suppressed the exciton quenching of (Cu-Li):NiCo$_2$O$_4$ SCS metal oxide, by passivating the surface defects (hydroxyl groups), as has been previously reported by other groups [23,27,39].

To investigate device performance effects that are based on the proposed bottom contact interface modification, OLED devices were fabricated using the configuration illustrated in Figure 3 to examine the impact of ITO/(Cu-Li):NiCo$_2$O$_4$/PEDOT:PSS on the device performance.

- Type A: ITO/PEDOT:PSS/SY/LiF/Al (reference device);
- Type B: ITO/(Cu-Li):NiCo$_2$O$_4$/SY/LiF/Al;
- Type C: ITO/(Cu-Li):NiCo$_2$O$_4$/PEDOT:PSS/SY/LiF/Al.

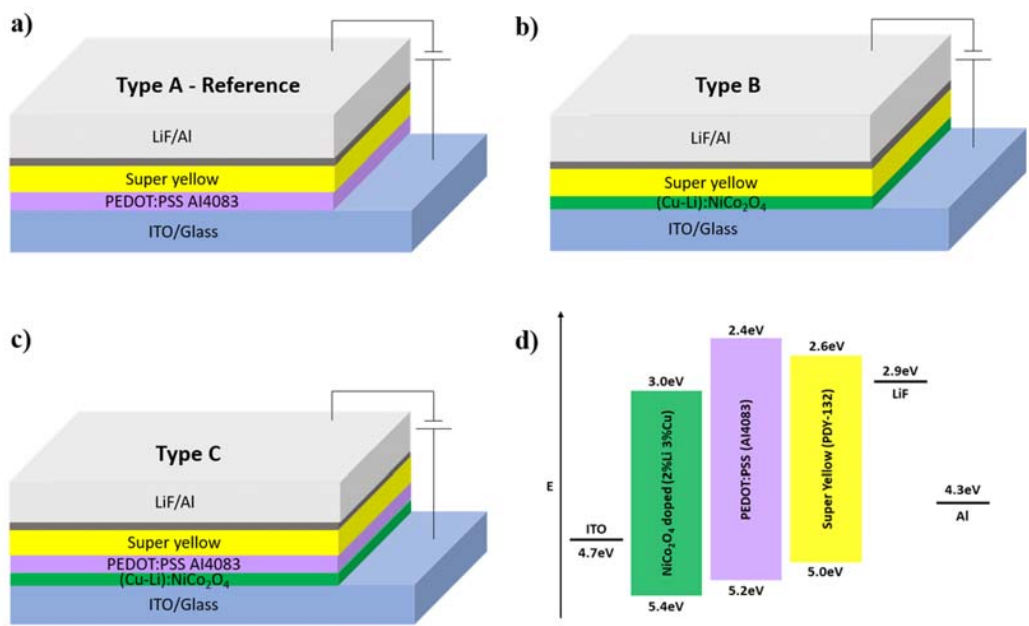

**Figure 3.** Schematic structures of the solution-processed OLED devices (**a–c**), (**d**) schematic of flat band diagram of the standard configuration super-yellow OLED device. Energy levels of each material were acquired from the following references, (Cu-Li):NiCo$_2$O$_4$ [28] PEDOT:PSS [40] and Super Yellow [41]**.**

Since ITO/(Cu-Li):NiCo$_2$O$_4$ was for the first time incorporated in OLED devices, a (Cu-Li):NiCo$_2$O$_4$ thickness optimization study was conducted and the acquired performance values are presented in Table S1 [Supporting Information]. The best OLED devices were obtained using a 15 nm (Cu-Li):NiCo$_2$O$_4$ film thickness, which therefore was used throughout the following experimental results.

Figure 4 shows the performance of the examined OLED devices, using the current density–voltage–luminance (JVL) characteristics. Detailed performance values, as well as statistical analysis, are also summarized in Table 1. As shown in Figure 4a, device type C clearly exhibits the highest current efficiency at every operating luminance. Moreover, as presented in Table 1, type C device delivers the highest current efficiency 5.6 cd/A and power efficiency efficacy 5.0 lm/W compared to both reference device type A (4.9 cd/A and 4.5 lm/W) and device type B (3.6 cd/A and 3.1 lm/W).

As shown in Figure 4b and Table 1, type B device shows increased diode threshold voltage and turn on voltage indicating the difficulty in charge injection. Since (Cu-Li):NiCo$_2$O$_4$ has a relatively high electrical conductivity and a smooth pinhole-free surface, then the difficulty in charge injection can be explained by improper energy level alignment. As illustrated in Figure 3d, the energy level barrier between the work functions of ITO and (Cu-Li):NiCo$_2$O$_4$ is 0.7 eV, which is higher compare to the 0.5 eV barrier between ITO and PEDOT:PSS, therefore resulting in an increased resistance of charge injection. Additionally, type B device presents two orders of magnitude higher leakage current compared to type A as shown in Figure 4c, and, in accordance with the finding of the PL measurements, strongly declines device performance. Finally, great dependency of electrical conductivity of SCS metal oxides on the device performance was also confirmed by investigating Cu:NiO$_x$ and CuO$_x$ instead of (Cu-Li):NiCo$_2$O$_4$, as presented in Figure S2 [Supporting information].

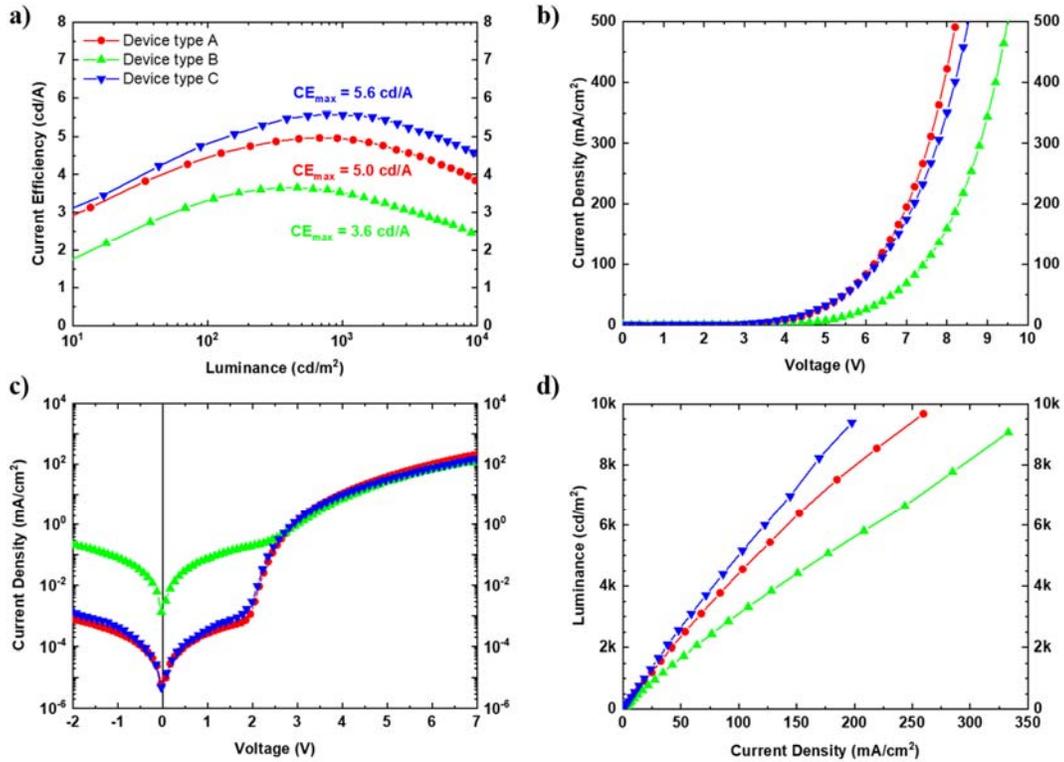

**Figure 4.** Electroluminescent performance of the different types of OLEDs devices, (**a**) current efficiency (J) as a function of luminance (L), (**b**) current density–voltage (JV) characteristics, (**c**) semi-log (JV) leakage current characteristics, and (**d**) luminance-current density (LJ) characteristics.

**Table 1.** Statistical analysis of the performance of OLED devices using different hole injection electrodes. Statistic sample size of 8 different devices were used to present average values, together with standard deviation, and maximum values in brackets.

| Device | Voltage (at 10 cd/m$^2$) [V] | Max. Luminance [cd/m$^2$] | Max. Current Efficiency [cd/A] | Max. Power Efficacy [lm/W] |
|---|---|---|---|---|
| Type A | 2.6 | 9.0 k ± 0.2 k | 4.9 ± 0.1 (5.0) | 4.4 ± 0.1 (4.5) |
| Type B | 2.8 | 9.0 k ± 0.5 k | 3.5 ± 0.1 (3.7) | 2.9 ± 0.2 (3.1) |
| Type C | 2.6 | 9.0 k ± 0.3 k | 5.4 ± 0.1 (5.6) | 4.8 ± 0.1 (5.0) |

The drawbacks of ITO/(Cu-Li):NiCo$_2$O$_4$ (type B) device was completely suppressed when PEDOT:PSS was incorporated within the device structure (device type C). As shown in Figure 4b,c, the type C device that is using ITO/(Cu-Li):NiCo$_2$O$_4$/PEDOT:PSS bottom electrode demonstrate almost the same JV and leakage current characteristics with the reference device type A using ITO/PEDOT:PSS bottom electrode. Additionally, the former (device type C) exhibit the same turn on voltage (see Table 1) and maximum luminance value (see Figure 2c) with the reference device (type A). Importantly, the device type C, which is based on the proposed bottom contact MO based interface modification, clearly demonstrated the highest luminance at the same current density across the entire scanned range, as it is presented in the luminance–current density plot (Figure 4d). Provided that the ITO/(Cu-Li):NiCo$_2$O$_4$/PEDOT:PSS bottom electrode has lower transmittance than the reference ITO/PEDOT:PSS bottom electrode (see Figure 2b), the improved luminance could not be attributed to light-outcoupling issues but mainly to the injection properties of the modified bottom electrode.

To explain the increased luminance, the investigated OLED devices were studied using hole only single-carrier device architecture. Hole only devices (HODs), with the structure of ITO/HIE/SY/MoO$_3$/Au with different hole injection electrodes, were implemented to ensure hole-only currents [42]. As shown in Figure 5a, in the forward bias regime, ITO/(Cu-Li):NiCo$_2$O$_4$/PEDOT:PSS/SY/MoO$_3$/Au shows the highest hole current density than the other hole-only devices in the entire voltage range. Since all the presented hole-only devices are using identical active layer the hole mobilities remain unaffected. The highest hole current density of the ITO/(Cu-Li):NiCo$_2$O$_4$/PEDOT:PSS bottom electrode-based hole-only devices indicates improved hole injection capability. The presented MOs within this study have a thickness of 15 nm; therefore, interfacial dipole effects are not expected to affect the energy levels and hole injection properties since such phenomena are observed in ultrathin layers [43,44]. Figure S3 [Supporting Information] shows vertical current density–voltage characteristics of the ITO/PEDOT:PSS, ITO/(Cu-Li):NiCo$_2$O$_4$ and ITO(Cu-Li):NiCo$_2$O$_4$/PEDOT:PSS bottom electrodes using Al as top contact. From the JV characteristics of the above device structures the highest current density observed in ITO/(Cu-Li):NiCo$_2$O$_4$/Al, followed by ITO/PEDOT:SS/Al and ITO/(Cu-Li):NiCo$_2$O$_4$/PEDOT:PSS/Al. Detailed investigation for the possibility of unintentional doping of the Ni-based MOs on PEDOT:PSS demands alternative advance characterization studies. However, the results presented within Figure S3 [Supporting Information] shown that ITO/(Cu-Li):NiCo$_2$O$_4$/PEDOT:PSS/Al resulted in lower vertical conductivity compared to ITO/PEDOT:PSS, indicating that unintentional doping effects are not expected to be the dominant factor for the improved device performance. According to the experimental results presented in this work, the improved hole injection is influenced by the wetting properties of the proposed MOs (Figure 1). Improved wetting influenced the morphology/topography of the overlayer PEDOT:PSS (Figure 2a) providing intimate interfaces and improving the hole injection properties of the bottom [ITO/(Cu-Li):NiCo$_2$O$_4$/PEDOT:PSS] electrode for the super-yellow OLED devices. To provide a direct confirmation of the improved hole injection capability of the proposed ITO/(Cu-Li):NiCo$_2$O$_4$/PEDOT:PSS bottom electrode the corresponding OLED devices were measured and compared, using the photocurrent mapping technique (PCT) [45]. As presented in Figure 5b, PCT images reveal significant differences in the respective OLED devices. The highest photocurrent values together with the most homogeneous photocurrent distribution were observed for device type C, incorporating the proposed ITO/(Cu-Li):NiCo$_2$O$_4$/PEDOT:PSS bottom electrode, followed by devices A (ITO/PEDOT:PSS bottom electrode) and B (ITO/(Cu-Li):NiCo$_2$O$_4$ bottom electrode) in agreement with the observations of hole-only device results and device efficiencies presented above. Further investigations of OLED devices with Cu:NiO$_x$, CuO$_x$, and un-doped NiCo$_2$O$_4$ interfacial layers used in alternative ITO/SCS metal oxide/PEDOT:PSS bottom electrode structure were performed, as presented in Figure S4 [Supporting Information]. As expected, in parallel to transparency properties that were discussed above (see Figure 2b) the electrical conductivity of the incorporated metal-oxide interfacial layer also plays a role on the performance of the bottom electrode. The best performing co-doped (Cu-Li):NiCo$_2$O$_4$ interfacial layer exhibited adequate transparency, the highest conductivity of ~4 S cm$^{-1}$ and the lowest static contact angle of ~6°. While, despite the relatively good wetting properties [~17°–Figure S1a], the lower electrical conductivity of Cu:NiO$_x$ (~1.25 × 10$^{-3}$ S cm$^{-1}$) [33] in comparison to (Cu-Li):NiCo$_2$O$_4$ and un-doped NiCo$_2$O$_4$ (~4 S cm$^{-1}$) [28] reduced the device performance. The worst performance was recorded in the device with CuO$_x$ due to the combination of poor wetting [~47°–Figure S1b] and low electrical conductivity ~1 × 10$^{-6}$ S cm$^{-1}$ [46]. Therefore, it can be concluded that the device performance of the proposed interface modification ITO/MO/PEDOT:PSS is depended not only on the common optoelectronic properties (transparency, conductivity, energy levels) of the bottom electrode but also on the wetting properties of the incorporated metal oxide interfacial layers that influence the coating of PEDOT:PSS overlayer and quality of device interfaces that resulted to OLEDs improve hole injection properties.

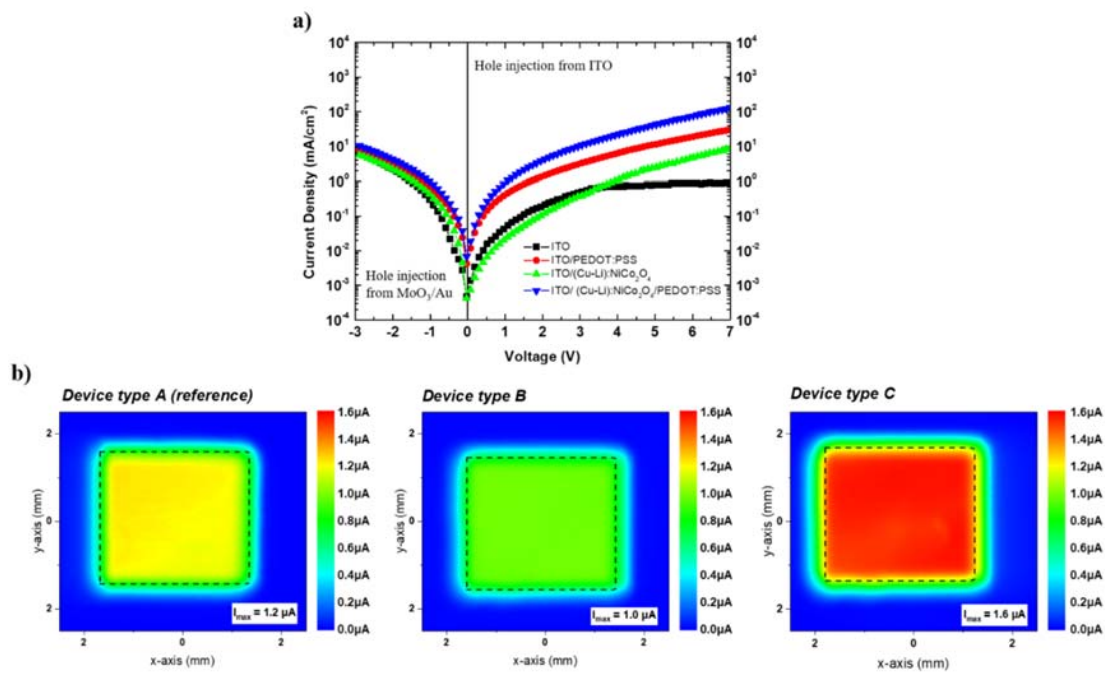

**Figure 5.** (**a**) JV characteristics of hole only devices with the structure of ITO/HIE/SY/MoO$_3$/Au, (**b**) photocurrent maps of complete dual carrier 3 × 3 mm$^2$ type A, B and C OLED devices.

## 4. Conclusions

The effect of appropriate solution-processed metal oxide interfacial layers between ITO and PEDOT:PSS on the performance of OLEDs was investigated. Improved wetting properties were achieved by using solution combustion synthesized Ni-based metal oxides [Cu:NiO$_x$, undoped-NiCo$_2$O$_4$ and co-doped (Cu-Li):NiCo$_2$O$_4$] but not with SCS CuO$_x$ interface modification. The improved performance of Ni-based metal oxides indicated the influence of Ni/oxygen vacancies on the presented wetting properties. We show that a range of Ni-containing solution-processed MO interfacial layers between ITO and PEDOT:PSS can be used to improve the bottom electrode contact and OLED device efficiency. The OLED device performance based on ITO/Ni-containing MOs/PEDOT:PSS bottom electrode was strongly correlated not only with the common optoelectronic properties but also with the MOs wetting properties. The best performing co-doped (Cu-Li):NiCo$_2$O$_4$ interfacial layer exhibited adequate transparency, the highest conductivity of ~4 S cm$^{-1}$ and the lowest static contact angle of ~6°. The wetting properties of co-doped (Cu-Li):NiCo$_2$O$_4$ resulted in a smooth and homogeneous PEDOT:PSS overlayer that provided intimate interfaces and improved the hole injection properties of the bottom electrode [ITO/(Cu-Li):NiCo$_2$O$_4$/PEDOT:PSS] for the super-yellow OLED devices, as experimentally verified by the presented photocurrent mapping and single carrier device studies. The incorporation of co-doped NiCo$_2$O$_4$ [(Cu-Li):NiCo$_2$O$_4$] as interfacial layer between ITO and PEDOT:PSS provided a high performance ITO/(Cu-Li):NiCo$_2$O$_4$/PEDOT:PSS bottom electrode that increased the super yellow OLEDs current efficiency and power efficacy by 12% and 11%, respectively. We believe that the proposed metal oxide-based interface modification can be used as a method to control wetting properties and to improve the bottom contact in a range of solution-processed light emitting diodes.

**Supplementary Materials:** The following are available online at www.mdpi.com/xxx/s1, **Figure S1**: Static contact angle of water droplet images on SCS metal oxide substrates, (a) ITO/Cu:NiO$_x$ and (b) ITO/CuO$_x$, **Table S1**: Statistical analysis of electroluminescence performance of lab-scale 9 mm$^2$ OLED devices investigating the effect of (Cu-Li):NiCo$_2$O$_4$ film thickness. Statistic sample size of 8 different devices was used to present average values, together with standard deviation, and maximum values in brackets; **Figure S2**: Performance of OLED devices based on ITO/SCS metal-oxide bottom electrodes, using different metal oxide interlayers, a) current efficiency over device luminance, b) semi-log JV - leakage current characteristics; **Figure S 3**: Current density – voltage (JV) characteristics of PEDOT:PSS, (Cu-Li):NiCo$_2$O$_4$ and (Cu-Li):NiCo2O4/PEDOT:PSS, sandwiched between ITO anode and Al cathode electrodes; **Figure S4**: Performance of OLED devices based on ITO/SCS metal-oxide/ PEDOT:PSS bottom electrodes, using different metal oxide interlayers, a) current efficiency versus device luminance, b) photocurrent mapping images of complete 3 x 3 mm$^2$ SY OLED devices;

**Funding:** This research was funded by European Research Council, grant number 647311. The C.S was funded by the Royal Society, grant number: NF171117.

**Acknowledgments:** This project received funding from the European Research Council (ERC) under the European Union's Horizon 2020 research and innovation program (Grant Agreement No.647311) and further supported from internal Cyprus University of Technology funding to the Molecular Electronics and Photonics Research Unit. C.S and D.D.C.B thank the Royal Society for the provision of a Newton International Fellowship to C.S (Application number: NF171117) that funded her stay in Oxford.

# *Supplementary Information*
# Bottom Contact Metal Oxide Interface Modification Improving the Efficiency of Organic Light Emitting Diodes

**Sergey M. Pozov [1], Apostolos Ioakeimidis [1], Ioannis T. Papadas [1], Chen Sun [2], Alexandra Z. Chrusou [1], Donal D. C. Bradley [2,3] and Stelios A. Choulis [1,\*]**

[1] Department of Mechanical Engineering and Materials Science and Engineering, Molecular Electronics and Photonics Research Unit, Cyprus University of Technology, Limassol 3603, Cyprus; sergey.pozov@cut.ac.cy (S.M.P.); a.ioakeimidis@cut.ac.cy (A.I.); ioannis.papadas@cut.ac.cy (I.T.P.); az.chrusou@cut.ac.cy (A.Z.C.)

[2] Department of Physics, Clarendon Laboratory, University of Oxford, Parks Rd, Oxford OX1 3PU, UK; chen.sun@physics.ox.ac.uk (C.S.); donal.bradley@kaust.edu.sa (D.D.C.B.)

[3] Physical Science and Engineering Division, King Abdullah University of Science and Technology (KAUST), Thuwal 23955-6900, Saudi Arabia

\* Correspondence: stelios.choulis@cut.ac.cy

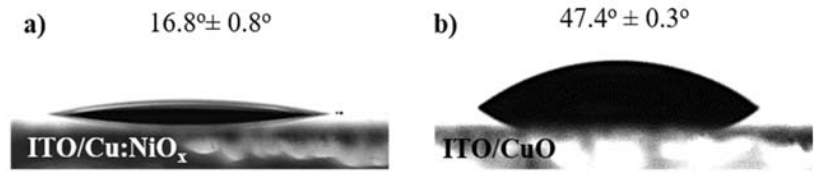

**Figure S1.** Static contact angle of deionized water droplet images on solution combustion synthesized (SCS) metal oxide substrates, a) ITO/Cu:NiO$_x$ and b) ITO/CuO$_x$.

**Table S1.** Statistical analysis of electroluminescence performance of lab-scale 9 mm$^2$ organic light emitting (OLED) devices investigating the effect of spinel film thickness. Statistic sample size of 8 different devices was used to present average values, together with standard deviation, and maximum values in brackets.

| Anode Electrode | Turn on voltage (at ~10 cd/m$^2$) [V] | Max. Luminance [cd/m$^2$] | Max. Current Efficiency [cd/A] | Max. Power Efficacy [lm/W] |
|---|---|---|---|---|
| ITO/PEDOT:PSS | 2.6 | 9.0k ± 0.2k | 4.9 ± 0.1 (5.0) | 4.4 ± 0.1 (4.5) |
| ITO/(Cu-Li):NiCo$_2$O$_4$ (~30 nm) | 2.7 | 9.0k ± 0.3k | 3.00 ± 0.4 (3.4) | 2.5 ± 0.5 (2.8) |
| ITO/(Cu-Li):NiCo$_2$O$_4$ (~20 nm) | 2.7 | 8.9k ± 0.2k | 3.2 ± 0.1 (3.4) | 2.5 ± 0.3 (2.9) |
| ITO/(Cu-Li):NiCo$_2$O$_4$ (~15 nm) | 2.7 | 9.0k ± 0.5k | 3.5 ± 0.1 (3.7) | 2.9 ± 0.2 (3.1) |
| ITO/(Cu-Li):NiCo$_2$O$_4$ (~10 nm) | 2.8 | 8.0k ± 1.5k | 3.4 ± 0.3 (3.7) | 2.8 ± 0.3 (3.0) |

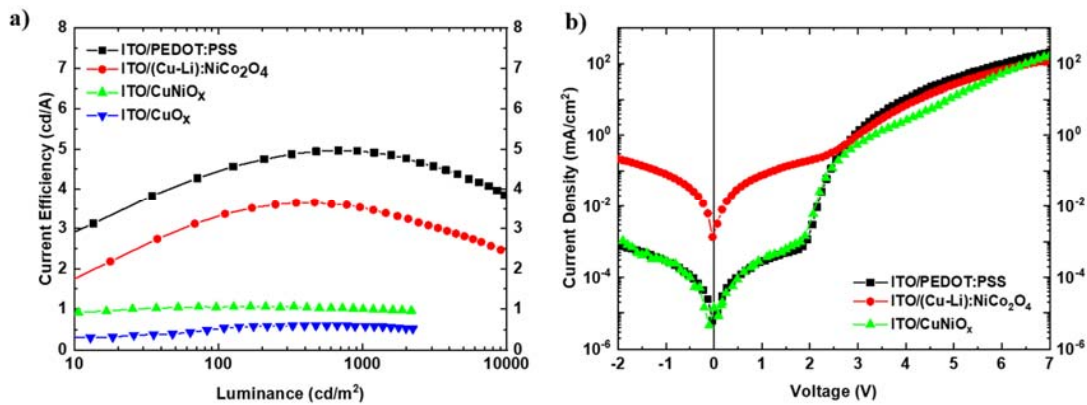

**Figure S1.** Performance of OLED devices based on ITO/SCS metal-oxide bottom electrodes, using different metal oxide interlayers, a) current efficiency over device luminance, b) semi-log *JV* - leakage current characteristics.

Replacing (Cu-Li):NiCo$_2$O$_4$ by Cu:NiO$_x$ or CuO$_x$ using a type B (indium tin oxide/metal oxide) device configuration, the performance analysis in Figure S2 points out to even lower efficiencies. Both Cu:NiO$_x$ and CuO$_x$ showed reduced performance in comparison to the optimized (Cu-Li):NiCo$_2$O$_4$, due to their much lower electrical conductivity Cu:NiO$_x$ (~1.25 x 10$^{-3}$ S/cm) [1] and CuO$_x$ (~1 x 10$^{-6}$ S/cm) [2], in comparison to (Cu-Li):NiCo$_2$O$_4$ (~4 S/cm) [3]. Studying the Figure S2b Cu:NiO$_x$ presents similar leakage current characteristics to our reference device ITO/PEDOT:PSS due to its high LUMO level (~1.8 eV) of [4] that ensures electron blocking from super-yellow active layer, however the much lower forward bias current lead to reduced device efficiency. The effect of proper energy level alignment and electrical conductivity of MO, is strongly distinct in CuO$_x$ with much deeper HOMO ~5.6 eV [5], strongly resisted the charge injection resulting in non-reproducible JVs, and significantly reduced device performance as shown in Figure S2a. Based on the fact that the low conductive PEDOT:PSS Al4083 formulation, used in this work has an electrical conductivity in the range of 10$^{-3}$ – 10$^{-4}$ S/cm, which is much lower that (Cu-Li):NiCo$_2$O$_4$ (~4 S/cm) we conclude that electrical conductivity plays a role among the examined MOs, however the proper energy level alignment with the active layer polymer is the most dominant parameter that affects the performance of OLED devices.

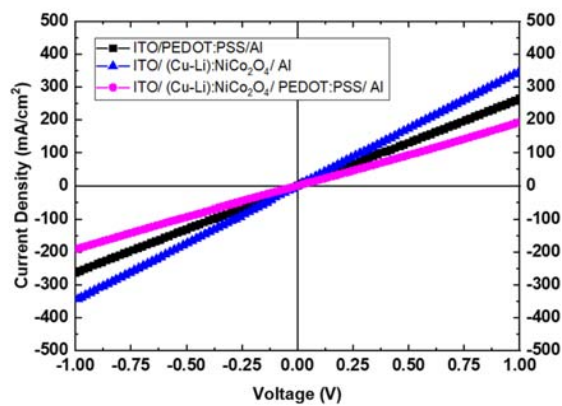

**Figure S 2.** Current density – voltage (JV) characteristics of PEDOT:PSS, (Cu-Li):NiCo$_2$O$_4$ and (Cu-Li):NiCo$_2$O$_4$/PEDOT:PSS, sandwiched between ITO anode and Al cathode electrodes.

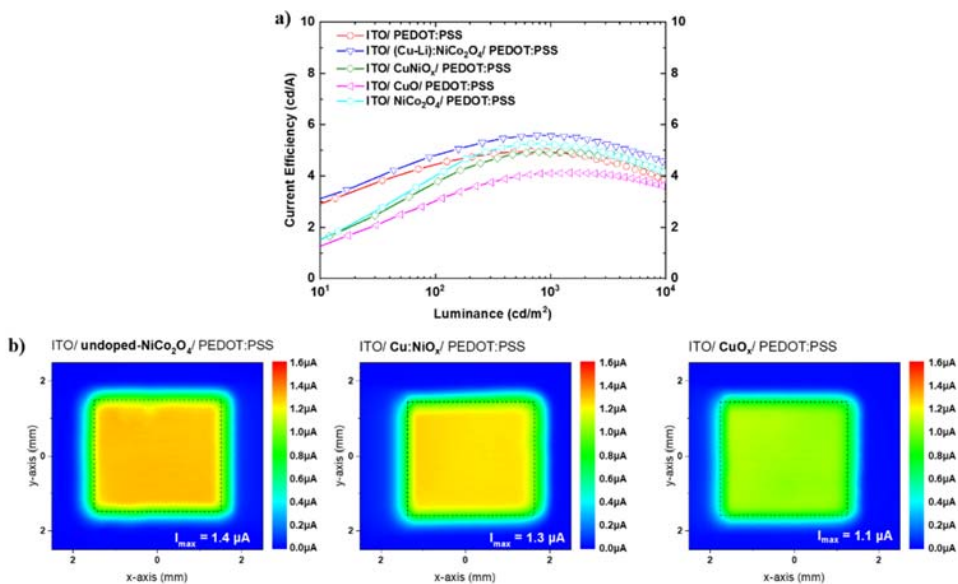

**Figure S4**. Performance of OLED devices based on ITO/SCS metal-oxide/ PEDOT:PSS bottom electrodes, using different metal oxide interlayers, a) current efficiency *versus* device luminance, b) photocurrent mapping images of complete 3 x 3 mm² SY OLED devices.

Device current efficiency presented in Figure S4a, follows the order, from highest to lowest, of ITO/undoped-NiCo$_2$O$_4$/PEDOT:PSS, followed by ITO/Cu:NiO$_X$/PEDOT:PSS and last ITO/CuO$_X$/PEDOT:PSS bottom electrode. The same trend was also observed in the generated photocurrent images Figure S4b, and the wetting properties of the examined SCS metal oxides (see Figure 1 and Figure S1).